\documentclass{jfm}

\usepackage{hyperref}
\hypersetup{
    colorlinks,
    linkcolor={red},
    citecolor={blue},
    urlcolor={blue}
}
\usepackage{graphicx}
\usepackage{epstopdf, epsfig}
\usepackage{amsmath}
\usepackage{fancyhdr}
\usepackage{graphicx}
\usepackage{caption}
\usepackage{subcaption}
\usepackage{mathrsfs}

\usepackage{wasysym}
\usepackage{soul}

\usepackage{tikz}
 
\definecolor{mycolor1}{rgb}{0.8027 0.2490  0.4430}

\shorttitle{Lagrangian time scale of rotation}
\shortauthor{Bordoloi and others}

\title{Lagrangian time scale of rotation for inertial fibers in isotropic turbulence}

\author{Ankur D. Bordoloi\aff{1}
  \corresp{\email{ankur.bordoloi@berkeley.edu}},
  Gautier Verhille\aff{2}, 
 \and Evan Variano\aff{1}}

\affiliation{\aff{1}Department of Civil and Environmental Engineering, University of California Berkeley, Berkeley, CA 94720, USA

\aff{2}Aix Marseille Univ, CNRS, Centrale Marseille, IRPHE, F-13013 Marseille, France

}

\begin{document}

\maketitle

\begin{abstract}
 Using time-resolved measurements of the orientation of rigid inertial fibers in a turbulence-tank, we investigate the {auto}correlation of their tumbling rate. The correlation time ($\tau_d$) is {well} predicted by {Kolmogorov} inertial-range scaling based on the fiber length ($L$) when the fiber inertia can be neglected. For inertial fibers, we propose a simple model considering fiber inertia {(measured by a tumbling Stokes number)} and a viscous torque which accurately predicts both  the correlation time and the variance of the tumbling rate. 

\end{abstract}
\section{Introduction}
Understanding the rotation of particles in turbulent flow is necessary for a wide range of scientific applications. For instance, the tumbling motion of fiber-like particles in turbulence plays a key role in the paper-making industry \citep{Lundell2011}. Rotation of planktonic organisms in oceanic turbulence is crucial for their locomotion, mating, forage, and escape from predators  \citep{Michalec2017}. Turbulent drag-reduction strategies based on introducing fibers into flow require understanding of fiber orientation and rotation \citep{MCCOMB1979, Sharma1980}.

{Several fundamental studies in the past few years have been devoted to the rotation of spheres~\citep{Zimmermann2011, Klein2013, Mathai2016}, of complex  particles~\citep{Pujara2018}, and axisymmetric anisotropic particles~(\citet{Voth2017} and reference therein.) }
Two important goals related to {these inquiries} are to understand: A) how the statistics of rotation relates to the particle size, and B) once set to motion by the ambient turbulence, how long a {particle} continues to preserve its rotation. The first piece of this puzzle has been addressed extensively in the past for both non-inertial (tracer) fibers \citep{Parsa2012, Pujara2017, Marchioli2013} and inertial fibers \citep{Parsa2014, Sabban2017,Kuperman2019, Bordoloi2017, Bounoua2018}. {Inertia might arise due to the density difference between the fiber and the fluid, and/or because of the fiber dimensions (length and diameter) being larger than the Kolmogorov length $\eta_{K}$}. In the latter category, long fibers with negligible diameter ($d \leq \eta_K$) were shown to obey Komogorov inertial-range scaling such that the variance of the tumbling rate, $\left<\dot{p}_i\dot{p}_i\right> \approx (L/\eta_K)^{-4/3}\tau_K^{-2}$ \citep{Parsa2014}. Here, $\eta_K$ and $\tau_K$ are Kolmogorov length and time scales, respectively. For fibers with large diameter ($d \gg \eta_K$) and small aspect ratio ($L/d$ = 1, 4), \citet{Bordoloi2017} modified this relationship by replacing particle length ($L$) by the spherical-volume-equivalent diameter $d_{eq} = (Ld^2)^{1/3}$. The $d_{eq}$-based scaling was also found to be consistent for particles of other complex shapes, such as cubes, cuboids and cones \citep{Pujara2018}. 

{In general, the rotation of a fiber can be described by the conservation of angular momentum}, written in the frame of the particle as
\begin{equation}
I{\dot{\mathbf{\Omega}}}+ \mathbf{\Omega} \times ( I \cdot {\mathbf{\Omega}} ) = \Gamma_f
\label{eq:01}
\end{equation}
\noindent
Here, $\mathbf{\Omega}$ is the total rotation (\emph{i.e.} spinning and tumbling) rate; $I$ is the moment of inertia of the fiber, and $\Gamma_f$ is the total torque applied on the fiber by the turbulent flow. Considering a viscous torque (linear in velocity {profile}), \citet{Bounoua2018} model the torque as $\Gamma_f \sim -4\pi\eta  \mathbf{\Omega} L^3/3 + 4\pi\eta\int \mathbf{u_f} \times \mathbf{s}ds$. {The first term corresponds to the viscous dissipation. The second term is the forcing term which is responsible for the transfer of energy from the fluid to the fiber.} The Coriolis term $\mathbf{\Omega} \times ( I \cdot {\mathbf{\Omega}} )$ {is generally neglected for long fibers assuming that the spinning rate is very small. This assumption seems justified {as the alignment of long fibers with a coarse grained vorticity is weak}~\citep{Pujara2019}}. The equation~\ref{eq:01} then reduces to a simplified Langevin equation,

\begin{equation}
%	\dot{{\Omega_t}} + \frac{1}{\tau_r}{\Omega_t} = \frac{1}{\tau_r}\xi.
\ddot{p}_i + \frac{1}{\tau_r}{\dot{p}_i} = \frac{1}{\tau_r}\xi.
\label{eq:02}
\end{equation}
\noindent
Here, {$\tau_r = I/4\pi\eta L^3$} is the rotational response time and {$\xi \sim \int \mathbf{u_f} \times \mathbf{s} ds/L^3$} is a colored noise related to the background turbulence. 
{The tumbling rate ($\dot{p}$) is then determined by the nature of $\xi$ and by the ratio of the response time of the particle ($\tau_r$) and the characteristic time  of the forcing $\tau_L\sim L/u_L \sim \epsilon^{-1/3}L^{2/3}$. This ratio defines the tumbling Stokes  number $St_{\dot{\mathbf{p}}}\sim \omega_L \tau_r$ which is equal to:}
\begin{equation}
St_{\dot{\mathbf{p}}} = \frac{1}{48}\frac{\rho}{\rho_f}\left( \frac{d}{\eta_K} \right) ^{4/3}\left(\frac{d}{L}\right)^{2/3}\left[1 + \frac{3}{4}\left( \frac{d}{L} \right)^2   \right].
\label{eq:stokes}
\end{equation}
{for a cylindrical fiber of length $L$ and diameter $d$.}

In a previous letter \citep{Bounoua2018}, we modeled $\xi$ as a Dirac function peaked at fiber length, $L$. This provided a theoretical basis to understand the influence of the fiber inertia on the variance of the tumbling rate via:
\begin{equation}
\left<\dot{p}_i\dot{p}_i\right> \sim \frac{1}{1+St_{\dot{\mathbf{p}}}^2}(L/\eta_K)^{-4/3}\tau_K^{-2}.
\label{eq:variance_gv}
\end{equation}
\noindent
The model stated in equation \ref{eq:variance_gv} unified results from \citet{Parsa2014}, \citet{Bordoloi2017}, and our experimental data over a wide range of aspect ratios. {This relation has also been verified recently by \citet{Kuperman2019} for long nylon fibers in air}.

% We revisit the f  Using slender-body approximations, \citet{MANSOO2005} computed the Lagrangian time scale of tumbling for fibers with negligible diameter and length upto 60$\eta_K$. They found a linear scaling (i.e. $\tau_d/\tau_K \approx L/\eta_K$) for fibers longer than 20$\eta_K$.

%The correlation time of the tumbling rate has been barely investigated. The main results come from numerical simulation. Shin and Koch showed that in the slender body approximation, the correlation time is constant for fiber length smaller than 10 eta_K and then increases with fiber length. For fiber smaller than the Kolmogorov length but heavier than the carrying fluid, Marchioli and Soldati showed that the correlation time increases with the Stokes number so with the fiber inertia. Extrapolating our model (Langevin equation+Dirac function for the forcing) to inertial fiber longer than the Kolmogorov length, we can derive that the correlation time scales with the forcing time scale so as $\tau_c\sim L^{2/3}$ and is always independent of the fiber inertia. This results is in contradiction with the results obtained for small fibers. We propose then to slightly modified our previous model to have a correct estimation of the correlation time.
 
{ While the variance of rotation rate has been investigated in detail, few studies have been devoted to the correlation time of rotation.}  The main results come from numerical simulation, which are limited to either short ($L\approx 10\eta_K$) \citep{Marchioli2013} or slender ($d<\eta_K$) \citep{MANSOO2005} fibers. \citet{MANSOO2005} showed that in the slender body limit, the correlation time is constant for fiber length smaller than $10\eta_K$ and then increases with fiber length. For fibers smaller than Kolmogorov length but heavier than the carrying fluid, \citet{Marchioli2013} showed that the correlation time increases with the Stokes number, {so with the fiber inertia. Extrapolating our previous model \citep{Bounoua2018} to inertial fibers longer than the Kolmogorov length, we find that the correlation time of the tumbling rate scales with the forcing time scale, $\tau_L\sim L^{2/3}$ independent of fiber inertia (see Section \ref{sc:results}). }
 
Herein, we take an experimental approach to measure the correlation time  of rotation for inertial fibers  over a wide range of length and diameter.  We also take the work of \citet{Bounoua2018} a step further and propose a new model that predicts both variance of tumbling rate and the {correlation} time scale of tumbling. In section \ref{sc:exp_method}, we review the experimental setup and describe our data analysis method. In Section \ref{sc:results}, we present our results and a theoretical model interpreting the results. Finally, we provide concluding remarks with a discussion about future research directions in section \ref{sc:conc}.
 
 % The experimental challenge resides in resolving long rotation trajectories, and thousands of them, to attain a statistical convergence. 
%This model was directly deduced from the 
%The model stated in equation \ref{eq:variance_gv} was re-examined by \citet{Kuperman2019} for rigid heavy fibers of length $l<17\eta_K$ in turbulent flow.

% In this communication,  to examine the timescale of rotational diffusion ($\tau_d$) of inertial fibers in homogeneous isotropic turbulence.  We also show that the statistical model proposed in \citet{Bounoua2018} is limited to the variance of rotation rate ($\left<\dot{p}_i\dot{p}_i\right>$) and fails in predicting $\tau_d$. In addition, we 

\section{Experimental setup and method}
\label{sc:exp_method}
\begin{table}
  \begin{center}
\def~{\hphantom{0}}
  \begin{tabular}{lccc}
Rotation frequency, $f$  & 5 Hz & 10 Hz & 15 Hz \\
Reynolds number, $Re_\lambda$  & 350 & 530 & 610 \\
Kolmogorov lengthscale, $\eta_K$ & 78.3 $\mu$m         &  46.6 $\mu$m       &       34.4 $\mu$m  \\ 
Kolmogorov timescale, $\tau_K$     &   6.14 ms      &        2.17 ms &     1.18 ms    \\
%Integral scale , $L_f$ \hl{GV: can you fill this}                               &         &         & \\
%Dissipation rate, $\epsilon$               &   0.03     &    0.21     &  0.72
  \end{tabular}
  \caption{Characteristics of turbulence in the cube tank facility. }
  \label{tab:1}
  \end{center}
\end{table}
The experimental setup consists of a cubic tank (each side = 60~cm) filled with water. We generate homogeneous and isotropic turbulence inside the tank by strategically stirring the water using 8 disks (diameter = 17~cm) with straight blades (height = 5~mm), each mounted on one corner of the tank. Each impeller is set to rotate independently via a 1.5 kW brushless motor at the same frequency but in the with opposite {chirality} to its adjacent three nearest neighbors. The turbulence inside the tank, set by the impeller frequency ($f$ = 5, 10 or 15 Hz), is characterized using standard PIV measurements \citep{Xu2013}, and is found to be fairly homogeneous and isotropic in the central region (volume $\approx$ 10 $\times$ 10 $\times$ 10 $\mathrm{cm^3}$). The relevant characteristics of turbulence in the tank are summarized in Table \ref{tab:1}.  Rigid polystyrene fibers cut to specific length ($L$ = 3.2 -- 40 mm) and diameter ($d$ = 0.5 - 2.5 mm) are introduced in dilute concentration ($<$ 0.01\% by volume) into the turbulence tank. We perform experiments on 18 different cases with aspect ratio ($\Lambda = L/d$) varying between 1.28 and 80. The details of each experimental case is provided in Table \ref{tab:2}. The density of polystyrene ($\rho_d$ = 1.04 $\mathrm{g/cm^3}$) makes the fibers near-neutrally buoyant, and the low fiber concentration allows to neglect the interaction between fibers and the retro action of the fiber on the turbulence.
\begin{table}
  \begin{center}
\def~{\hphantom{0}}
  \begin{tabular}{cccccc}
  Length  & Diameter  & Aspect ratio & Stokes number & Reynolds number & symbol\\
       ($L/\eta_K$) & ($d/\eta$) &  ($\Lambda$) & ($St_{\dot{p}}$) & ($Re_{\lambda}$) & \\\\
510.8 & 6.4 & 80 & 0.014 & 350 & $\color[rgb]{0.23031     0.01659     0.19198}\CIRCLE$\\
858.9 & 10.7 & 80 & 0.027 & 530 & $\color[rgb]{0.29475    0.037365     0.30885}\blacksquare$\\
255.4 & 8.2 & 31 & 0.035 & 350 & $\color[rgb]{0.33265    0.070431     0.43219}\CIRCLE$\\
127.7 & 8.2 & 15 & 0.056 & 350 & $\color[rgb]{0.34507     0.11698     0.55268}\CIRCLE$\\
429.5 & 13.7 & 31 & 0.071 & 530 & $\color[rgb]{0.33537     0.17689     0.66183}\blacksquare$\\
40.9 & 8.2 & 5 & 0.123 & 350 & $\color[rgb]{0.30882      0.2488     0.75259}\CIRCLE$\\
127.7 & 12.8 & 10 & 0.138 & 350 & $\color[rgb]{0.27205     0.33027     0.81991}\CIRCLE$\\
429.5 & 21.5 & 20 & 0.172 & 530 & $\color[rgb]{0.23243     0.41807     0.86108}\blacksquare$\\
81.7 & 12.8 & 6 & 0.187 & 350 & $\color[rgb]{0.19748     0.50841     0.87588}\CIRCLE$\\
255.4 & 25.5 & 10 & 0.347 & 350 & $\color[rgb]{0.17416     0.59734     0.86655}\CIRCLE$\\
137.4 & 21.5 & 6 & 0.375 & 530 & $\color[rgb]{0.1683     0.68105     0.83757}\blacksquare$\\
127.7 & 25.5 & 5 & 0.563 & 350 & $\color[rgb]{0.18411     0.75623     0.79527}\CIRCLE$\\
429.5 & 42.9 & 10 & 0.694 & 530 & $\color[rgb]{0.22376     0.82035     0.74727}\blacksquare$\\
51.1 & 20.4 & 2 & 0.722 & 350 & $\color[rgb]{0.28722     0.87188     0.70183}\CIRCLE$\\
127.7 & 31.9 & 4 & 0.894 & 350 & $\color[rgb]{0.3722     0.91047     0.66714}\CIRCLE$\\
214.7 & 42.9 & 5 & 1.126 & 530 & $\color[rgb]{0.47423       0.937      0.6506}\blacksquare$\\
85.9 & 34.4 & 2 & 1.443 & 530 & $\color[rgb]{0.58705     0.95356     0.65815}\blacksquare$\\
214.7 & 53.7 & 4 & 1.788 & 530 & $\color[rgb]{0.703     0.96331     0.69368}\blacksquare$\\
291 & 72.8 & 4 & 2.682 & 610 & $\color[rgb]{0.81365     0.97028     0.75856}\blacktriangle$\\
  \end{tabular}
  \caption{Fiber-dimensions ($L,d$) normalized by Kolmogorov length scale ($\eta_K$), aspect ratio ($\Lambda = L/d$), and rotational Stokes number ($St_{\dot{p}}$) of fibers tested under three specific Reynolds number ($Re_{\lambda}$) of the background turbulence. Each case is specified by a symbol color-coded in the ascending order of $St_{\dot{p}}$ . The three $Re_{\lambda}$ (see Table \ref{tab:1}) are designated by the symbol shape. }
  \label{tab:2}
  \end{center}
\end{table}

We image the fibers on {two} orthogonally arranged 1-MP-high-speed-cameras, all cameras being triggered  simultaneously at a frame rate of 0.5-1 kHz. The fibers are backlit onto each camera by an accompanying LED panel with diffuser. Each fiber is reconstructed into the 3D space using a custom MATLAB code. First, the extremeties of each fiber are detected in all {two} images. Then, those extremeties are triangulated into the 3D space. Finally, the centroid location ($x_{c,i}$) and the orientation ($p_i$) of each fiber are optimized such that the difference between the projection of the fiber onto each camera and the actual image is minimized. The measurement volume is that of a cube of $\approx$ 13 cm length centered at the center of the tank. $N \geq 5000$ individual trajectories are stored and used to compute the rotation statistics  for each case.

{The characteristic time of the dynamics of a random signal is given by the correlation function. The correlation function of $\dot{p}_i(t)$ is defined as,}
\begin{equation}
C_{\dot{p}_i}(\tau) = \frac{ \left< \dot{p}_i(\tau)\dot{p}_i(t+\tau)\right>}{\left< \dot{p}_i\dot{p}_i\right>}.
\label{eq:corr1}
\end{equation}
\noindent
{Here there is no summation over $i$.} For our calculations, we consider trajectories which are longer than 10 ms. To avoid bias in  the mean of $C_{\dot{p}_i}$ due to {the correlation between trajectory length and particle speed}, we compute the mean weighted on trajectory length, such that 
\begin{equation}
\bar{C}_{\dot{p}_i} (\tau)= \frac{\sum_{k=1}^NC_{\dot{p}_i}(\tau) T_k}{\sum_{k=1}^N T_k},
\label{eq:corr2}
\end{equation} 
\noindent
where $T_k$ is the length of a trajectory. For our analysis, we use the average of the three components of $\bar{C}_{\dot{p}_i}$, which were statistically indistinguishable. We denote the average as $\bar{C}_{\dot{p}} (\tau)$ and use it to compute two time scales of rotation. The first time scale is based on the zero-crossing time ($\tau_{d1}$) of the Lagrangian autocorrelation function $\bar{C}_{\dot{p}} (\tau)$ as described in \citet{MANSOO2005}. The second time scale is the integral time scale computed as:
\begin{equation}
\tau_{d2} = \int_0^\infty\bar{C}_{\dot{p}} (\tau)^2d\tau \approx \sum \bar{C}_{\dot{p}} (\tau)^2 \Delta \tau    
\end{equation}
{which is more tractable theoretically as shown later.} 

\begin{figure}
\centering

	 \begin{subfigure}[b]{0.4\textwidth}
        \includegraphics[width=\textwidth]{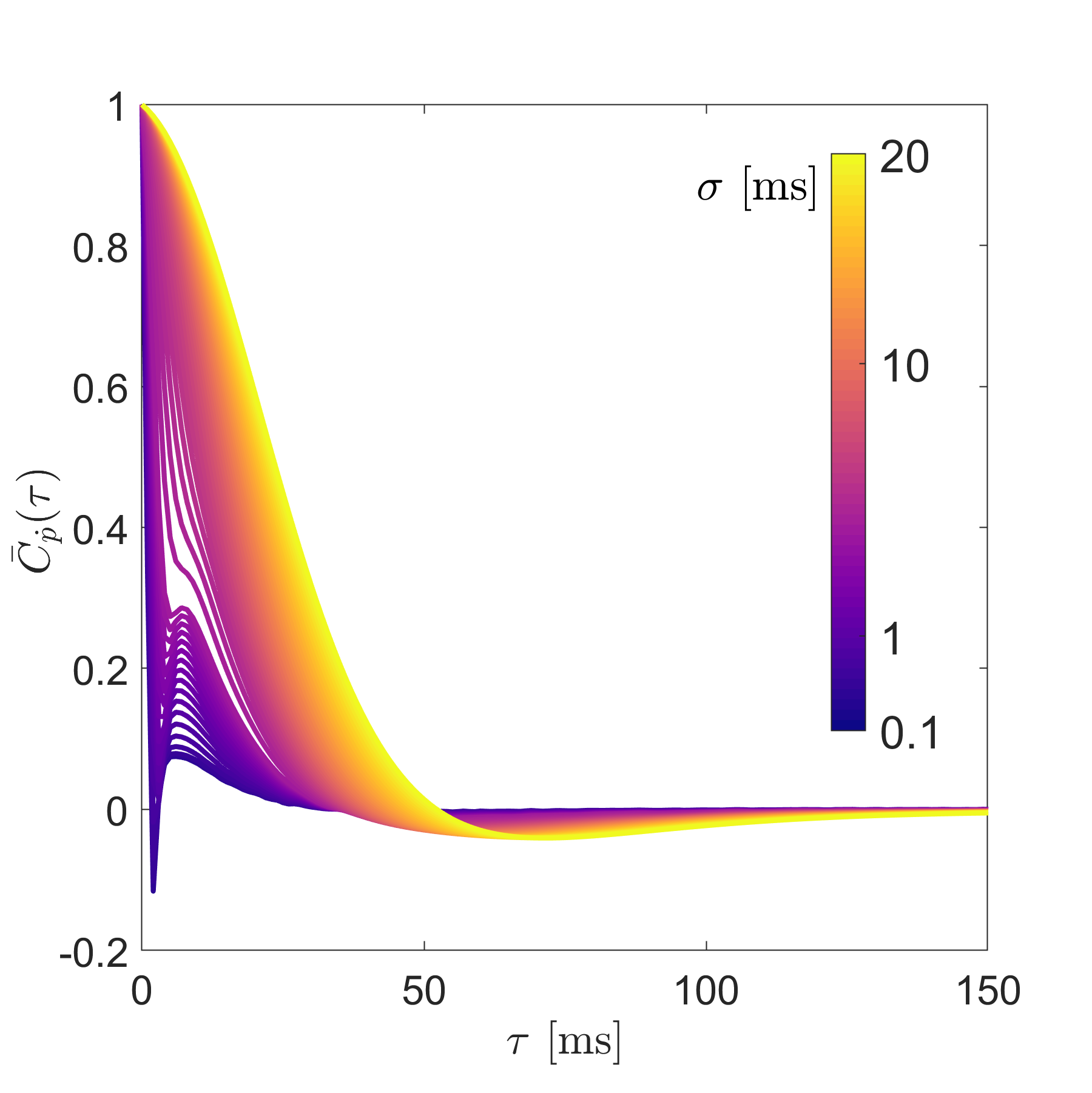} 
         \caption{}
        \label{fig:autocor_evol}
        \end{subfigure}
        %\vspace{20pt}
    \begin{subfigure}[b]{0.4\textwidth}
        \includegraphics[width=\textwidth]{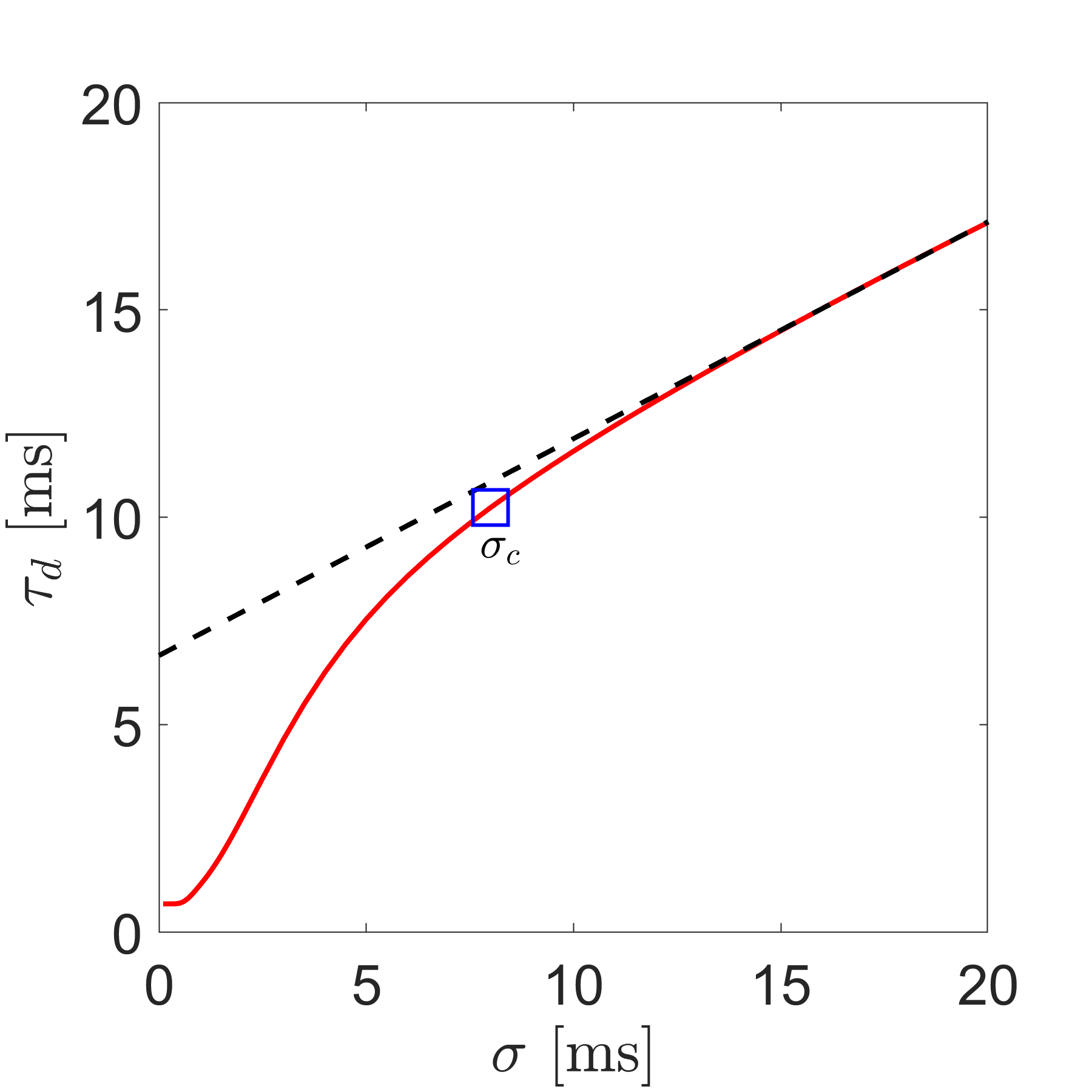} 
         \caption{}
        \label{fig:tauc_evol}
        \end{subfigure}

 \caption{a) Lagrangian autocorrelation of fiber tumbling rate, and b) the integral time scale of rotation ($\tau_{d2}$) of a representative fiber  for different filtering window size.}\label{fig:filter}
\end{figure}

Before computing the statistics of rotation, the experimental noise in $p_i(t)$ is removed by filtering it through a series of one-dimensional Gaussian kernels of window-size $\sigma \le 20$~ms \citep{Mordant2004, Volk2007}. The tumbling rate ($\dot{p}_i(t)$) for each $\sigma$ is then computed using a symmetric second-order central-difference scheme.  Figure \ref{fig:filter} demonstrates this method by showing the effect of $\sigma$ on {(a)}: the mean autocorrelation of tumbling rate $\bar{C}_{\dot{p}}(\tau)$, and {(b)}: the integral time-scale of rotational dispersion ($\tau_{d2}$). Assuming that the experimentally measured $\dot{p}_i(t)$ contains only uncorrelated noise, we extract the noise-free  $\tau_{d2}$ by fitting a straight line for the linear segment ($\sigma \ge 10$ in this example) of each plot and extrapolating it to $\sigma =0$ (see figure \ref{fig:tauc_evol}). We use a critical $\sigma_c$  to compute the noise-free mean autocorrelation ($\bar{C}_{\dot{p}} (\tau)$) of rotation and the zero-crossing time ($\tau_{d1}$). We choose $\sigma_c$ to be the smallest $\sigma$ at which the filtered data deviated from the fit by less than 10\%. {We tested the sensitivity of this criterion by varying it between 5-20\% and did not find it to affect our results.}

\section{{Correlation time scale of tumbling rate}}
\label{sc:results}
\subsection{Experimental observation}

\begin{figure}
\centering
	 \begin{subfigure}[b]{0.4\textwidth}
        \includegraphics[width=\textwidth]{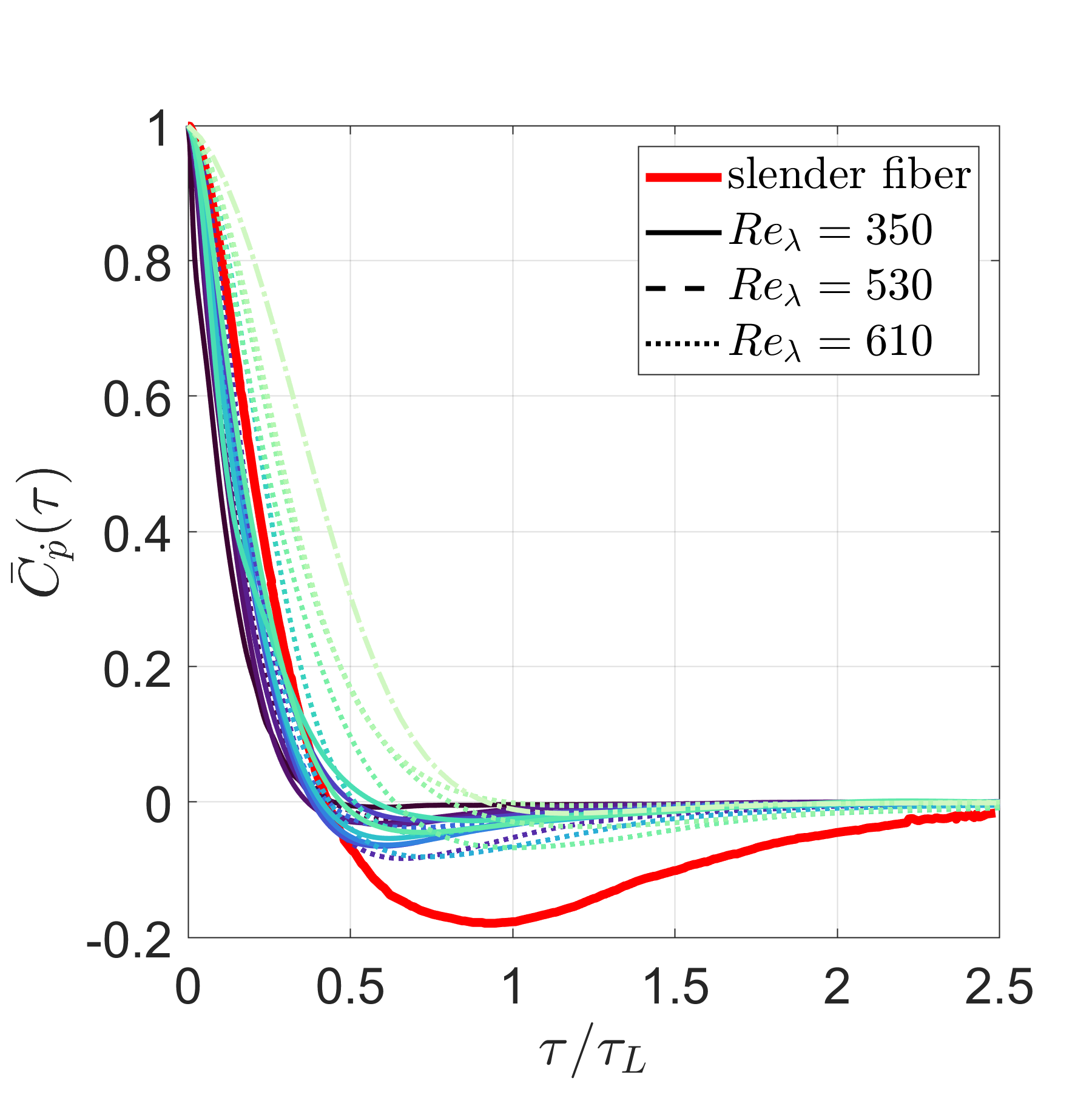} 
         \caption{}
        \label{fig:autocorr_norm}
        \end{subfigure}
          %\vspace{20pt}
    \begin{subfigure}[b]{0.4\textwidth}
        \includegraphics[width=\textwidth]{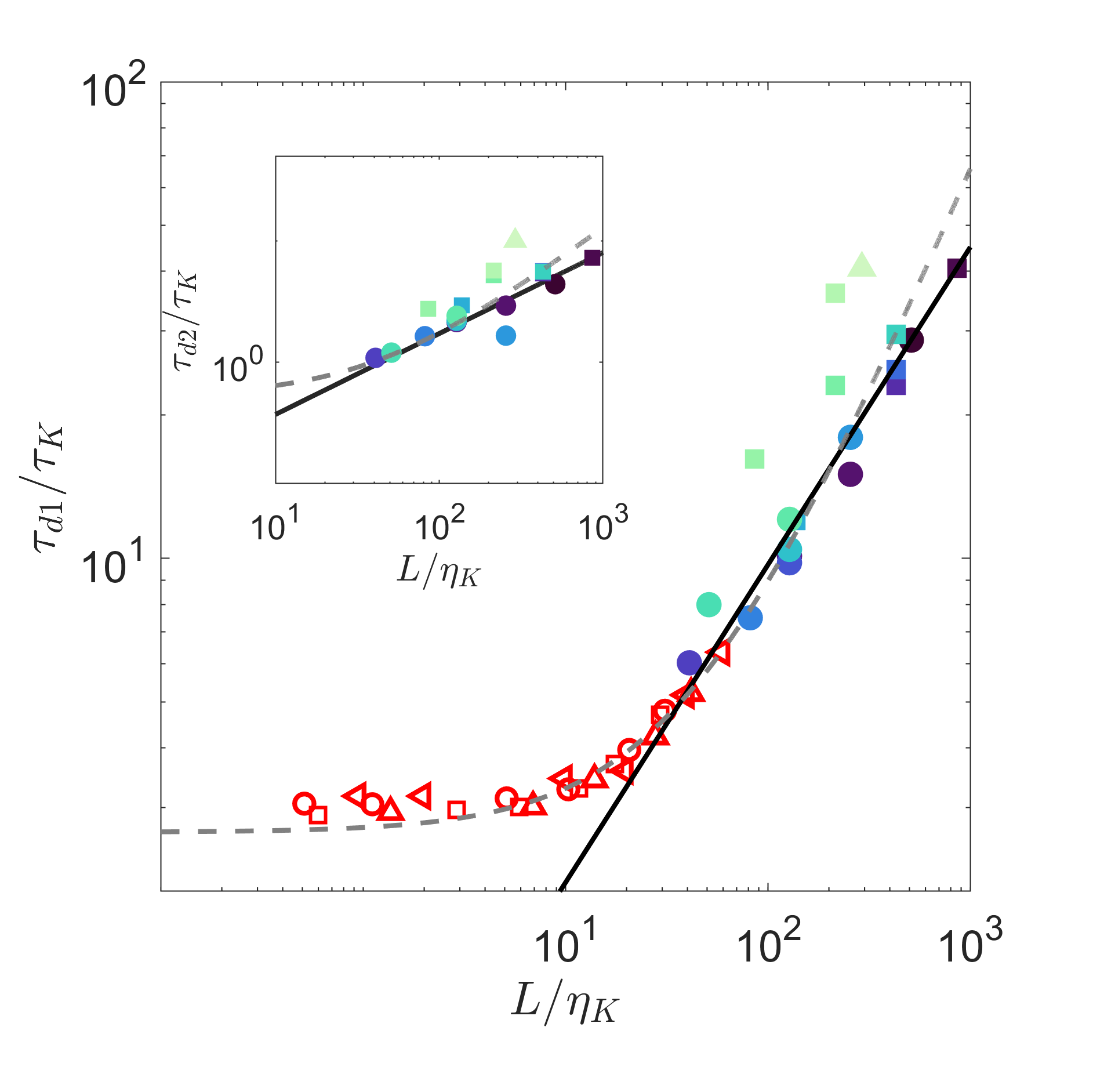} 
         \caption{}
        \label{fig:corr_time}
        \end{subfigure}
        
        %  \begin{subfigure}[b]{0.65\textwidth}
        % \includegraphics[width=\textwidth]{legend_CH.png} 
        %  %\caption{}
        % \label{}
       % \end{subfigure}

 \caption{a) Lagrangian autocorrelation $\bar{C}_{\dot{p}} (\tau)$ of tumbling rate,  b) the zero-crossing time $\tau_{d1}$  and the integral time scale $\tau_{d2}$ (see inset) for inertial fibers. Numerical result from \citet{MANSOO2005} for a slender fiber with $L=41.7\eta_K$, $Re_{\lambda}=39.9$ is included in a) and slender fibers at multiple lengths and $Re_{\lambda}$ = 16.5 (\textcolor{red}{$\square$}), 30.7 (\textcolor{red}{$\Circle$}), 39.9 (\textcolor{red}{$\triangle$}) and 53.3 (\textcolor{red}{$\triangledown$}) are included in (b)  for comparison. The dashed lines in (b) represent the linear fit ($\tau_d/\tau_K = 2.66 + 0.063L/\eta_K$) suggested in \citet{MANSOO2005}.  This fit is multiplied by the mean ratio of the zero crossing time and the integral time from our data in the inset. Data include particles from very little inertia ($\color[rgb]{0.23031     0.01659     0.19198}\CIRCLE$) to those with $St_{\dot{p}} > 0.7$ ($\color[rgb]{0.28722     0.87188     0.70183}\CIRCLE$ $\color[rgb]{0.3722     0.91047     0.66714}\CIRCLE$  $\color[rgb]{0.47423       0.937      0.6506}\blacksquare$  $\color[rgb]{0.58705     0.95356     0.65815}\blacksquare$  $\color[rgb]{0.703     0.96331     0.69368}\blacksquare$  $\color[rgb]{0.81365     0.97028     0.75856}\blacktriangle$).}\label{fig:kolm_corr}
\end{figure}

{If we neglect inertia and assume that a fiber {of size $L$} is rotated only by eddies of size $L$, the tumbling rate of the fiber should correlate to a timescale $\tau_L \sim L/u_L \approx (L/\eta_K)^{2/3}\tau_K$, where $u_L$ is the typical velocity at scale $L$.} The mean Lagrangian autocorrelation {function} $\bar{C}_{\dot{p}} (\tau)$ of tumbling for various fibers {are shown in figure \ref{fig:autocorr_norm}. This plot includes our measurements and the longest fiber ($L = 41.7\eta_K$) simulated by \citet{MANSOO2005} at $R_{\lambda}$ = 39.9 neglecting fiber inertia ($I=0$)}.  {With the horizontal-axis normalized by $\tau_L$, the measurements of the autocorrelation function for fibers with $St_{\dot{p}} < 0.7$ are  independent of $St_{\dot{p}}$ and close to the one obtained by \citet{MANSOO2005}}. In all these cases, the  zero-crossing time is $\tau_{d1}$  =  0.43$\tau_L$ with a 95\% confidence interval  $\pm 0.02$. The difference from the slender body {approximation} appears after the zero-crossing time, such that  our measurements decorrelate on a shorter time scale than simulations by \citep{MANSOO2005}. For fibers with $St_{\dot{p}} > 0.7$,  $\bar{C}_{\dot{p}} (\tau)$ becomes sensitive to $St_{\dot{p}}$, such that the zero-crossing time increases with $St_{\dot{p}}$.

%larger Kolmogorov's inertial scaling works on the autocorrelation function for fibers with Stokes number   tracer fiber from .  In figure \ref{fig:autocorr_evol}

In figure \ref{fig:corr_time}, we directly show the evolution of zero-crossing time ($\tau_{d1}$) normalized by the Kolmogorov time scale ($\tau_K$) with respect to normalized fiber length ($L/\eta_K$). We also include the zero-crossing time for all slender  fibers  computed by \citet{MANSOO2005}. Irrespective of the Reynolds number ($Re_{\lambda}$), a 2/3 power-law scaling qualitatively captures the evolution of $\tau_{d1}$ for fibers with $St_{\dot{p}}<0.7$. {Also, the zero-crossing times reported in \citet{MANSOO2005} approach  this power-law scaling as their length enters the inertial range.} A similar plot for the integral time ($\tau_{d2}$) is shown in inset that also demonstrates the 2/3 power-law fit. \citet{MANSOO2005} {proposed a linear fit for $\tau_{d1}$ from their simulations with long fibers ($25\eta < L < 60\eta$); the inertial-range scaling was not obvious there  because of the limited range of fiber length they simulated. }

Our data agree with the scaling law when $St_{\dot{p}}<0.7$, but not when $St_{\dot{p}}>0.7$. To investigate this effect,  we propose an improved version of the model proposed in \citet{Bounoua2018} which captures both the evolution of the variance and the tumbling rate for inertial fibers ($L\geq10\eta_K$).

\subsection{Theoretical model}
In \citet{Bounoua2018}, we modeled the forcing torque $\xi$ as a Dirac function peaked at the fiber length, $L$. As we saw, this assumption fails to predict the effect of fiber inertia on the correlation time scale ($\tau_d$) measured from our experiments. {Here, we assume that the process of filtering due to the integration of the viscous forces along the fiber length is smoother and can be described by a bandpass filter peaked on $\omega_L\sim \epsilon^{-1/3}L^{2/3}$. In that case, the forcing torque $\xi$ in Fourier space can be written as,
\begin{equation}
\xi(x) = \xi_L \frac{1}{1 + \imath Q(x-1/x)},
\label{eq:03}
\end{equation}}

\begin{figure}
\centering
	 \begin{subfigure}[b]{0.3\textwidth}
        \includegraphics[width=\textwidth]{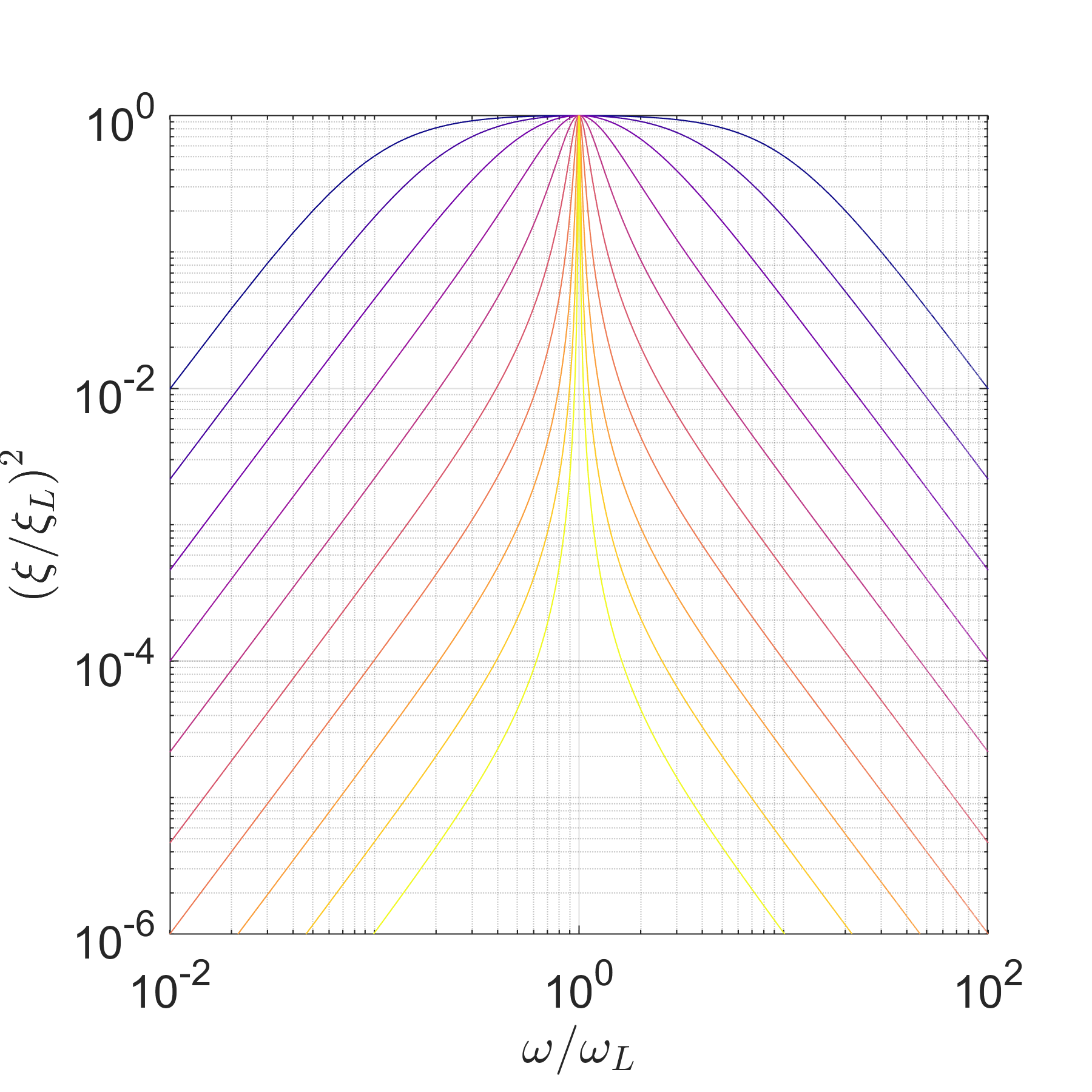} 
         \caption{}
        \label{fig:spec_response} 
        \end{subfigure}
	 \begin{subfigure}[b]{0.3\textwidth}
        \includegraphics[width=\textwidth]{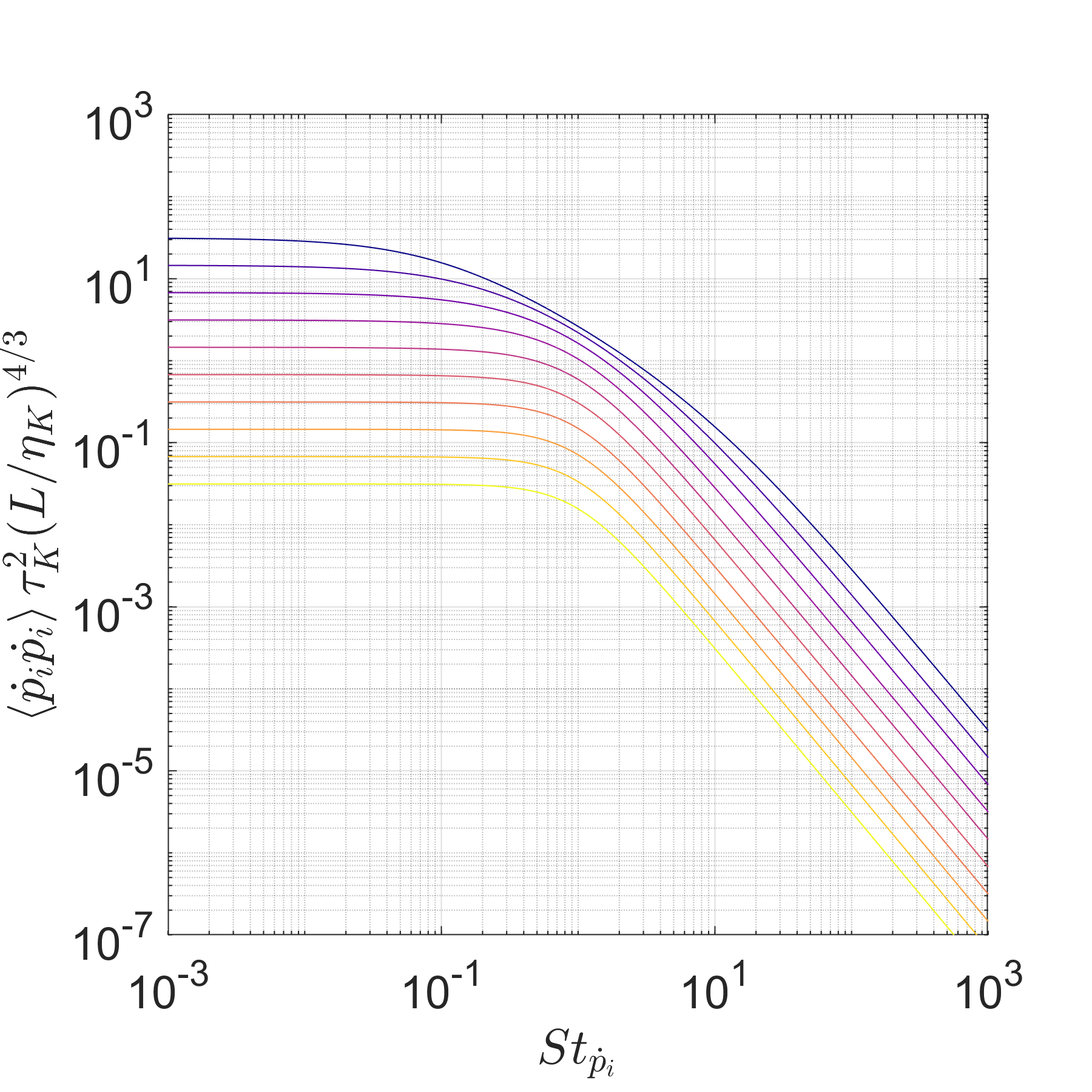} 
         \caption{}
        \label{fig:Qsigma}
        \end{subfigure}
        %\vspace{20pt}
    \begin{subfigure}[b]{0.3\textwidth}
        \includegraphics[width=\textwidth]{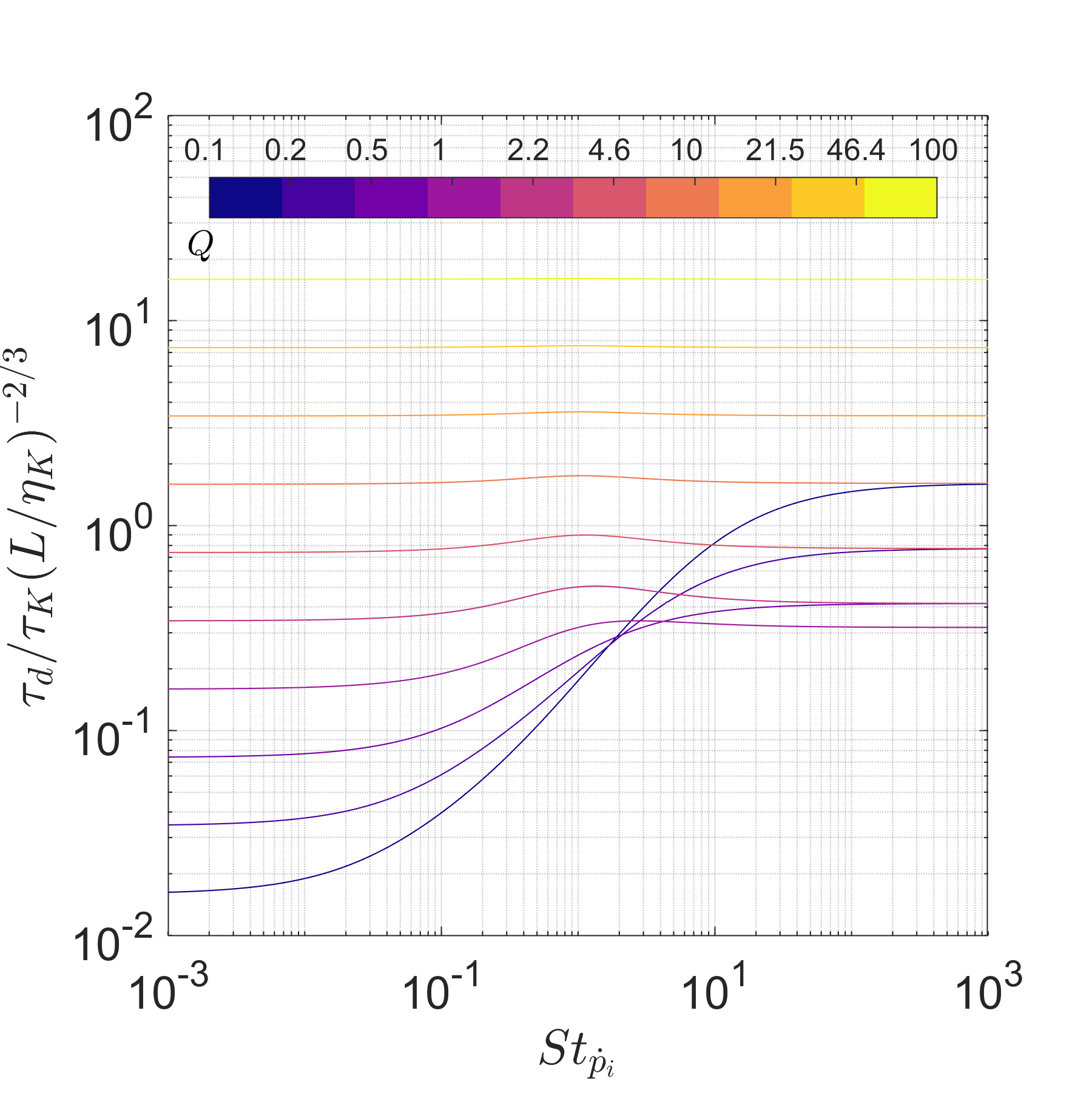} 
         \caption{}
        \label{fig:Qtau}
        \end{subfigure}

 \caption{The effect of quality factor ($Q$) on  a) the white noise spectrum, the transfer functions for b) variance of tumbling rate and c) correlation time scale. The colormap indicates the value of $Q$ used for this computation.}
\end{figure}

\noindent
{with $x=\omega/\omega_L$ and $Q$ is the quality factor of the filter}. {$\xi_L$ is the amplitude of the turbulent spectrum at scale $L$, such that {$|\xi_L|^2 \sim \tau_K^{-2}(\eta_K/L)^{4/3}$}.  The quality factor $Q$  determines the width of the band-pass filter as shown on figure~\ref{fig:spec_response}. }The spectrum reduces to the Dirac function  when $Q \rightarrow \infty$. 

The solution of equation \ref{eq:02} leads to:
\begin{equation}
    \dot{p}_i = \frac{\xi(\omega)}{1+\imath\omega\tau_r}.
    \label{eq:gv}
\end{equation}
For a given spectrum of $\xi_L$, one can determine the variance and the correlation time of tumbling rate from equation \ref{eq:gv}. For simplicity, we will assume that $\xi_L$ is a white noise to derive analytical expression for the variance and the tumbling rate. This assumption should hold as long as the quality factor is not too low and that the spectrum is indeed peaked at the frequency $\omega_L$. Further, { this assumption will be justified by the agreement between the model and our experimental results (which do not match when the Dirac function selects the amplitude of the spectrum only at the frequency $\omega_L$)}. Within this framework, the variance of $\dot{p}_i$ is:

\begin{equation}
\left\langle \dot{p}_i\dot{p}_i\right\rangle = \xi_L^2\int \frac{1}{1+ St_{\dot{p}} ^2x^2}\frac{x^2}{x^2 + Q^2(x^2-1)^2}dx.
\label{eq:04}
\end{equation} 
\noindent
{In a similar vein, we can derive an analytical expression for the correlation time ($\tau_d$) of the tumbling rate:}
\begin{equation}
\tau_d= \int C_{\dot{p}}(t)^2dt = \int \hat{C}_{\dot{p}}(\omega)\hat{C}_{\dot{p}}^{*}(\omega)d\omega,
\label{eq:05}
\end{equation}
\noindent
{where, $\hat{C}_{\dot{p}}(\omega)$ is the Fourier transform of the autocorrelation of $\dot{p}_i$. Contrary to the definition of the zero-crossing time, this expression is suitable analytically.  From equation~\ref{eq:03} and~\ref{eq:gv}, the modulus of the correlation function $\hat{C}$ can be written:}

\begin{equation}
\left|\hat{C}_{\dot{p}}(x)\right|^2 =  \frac{1}{\left<\dot{p}_i\dot{p}_i\right>^2} \left( \frac{\xi_L^4}{1+St^2 x^2} \frac{x^4}{(x^2+Q^2(x^2-1)^2)^2}\right).
\label{eq:dum}
\end{equation}

\noindent
Hence, the correlation time $\tau_d$ is,
\begin{equation}
\tau_d = \frac{\xi_L^4}{\left<\dot{p}_i\dot{p}_i\right>^2}\int \frac{\beta}{1+\alpha St_{\dot{p}}^2x^2}\frac{x^4}{(x^2 + Q^2(x^2-1)^2)^2}dx.
\label{eq:07}
\end{equation}

The solutions for the two transfer functions: $\left<\dot{p}_i\dot{p}_i\right>\tau_K^{-2}(L/\eta_K)^{4/3}$ and $\tau_d(L/\eta_K)^{-2/3}/\tau_K$ are shown for multiple $Q$ values in figure \ref{fig:Qsigma} and \ref{fig:Qtau}, respectively. The rescaled {variance $\left<\dot{p}_i\dot{p}_i\right>\tau_K^{2}(L/\eta_K)^{-4/3}$} for all $Q$ values show similar trends with a plateau for low Stokes number, and its value decreasing as $St_{\dot{p}}^{-2}$ for higher Stokes numbers as observed for the Dirac function approximation in \citet{Bounoua2018}. Also, the Stokes number where the transition between these two regimes occurs increases with decreasing $Q$. The correlation time for the large values of $Q$ is almost constant irrespective of Stokes number, which is a trend observed for the Dirac function formulation in \citet{Bounoua2018}.  For smaller values of $Q$, we observe that the correlation time increases beyond $St_{\dot{p}} \approx 1$ and eventually reaches a plateau. 

\begin{figure}
\centering
	 \begin{subfigure}[b]{0.45\textwidth}
        \includegraphics[width=\textwidth]{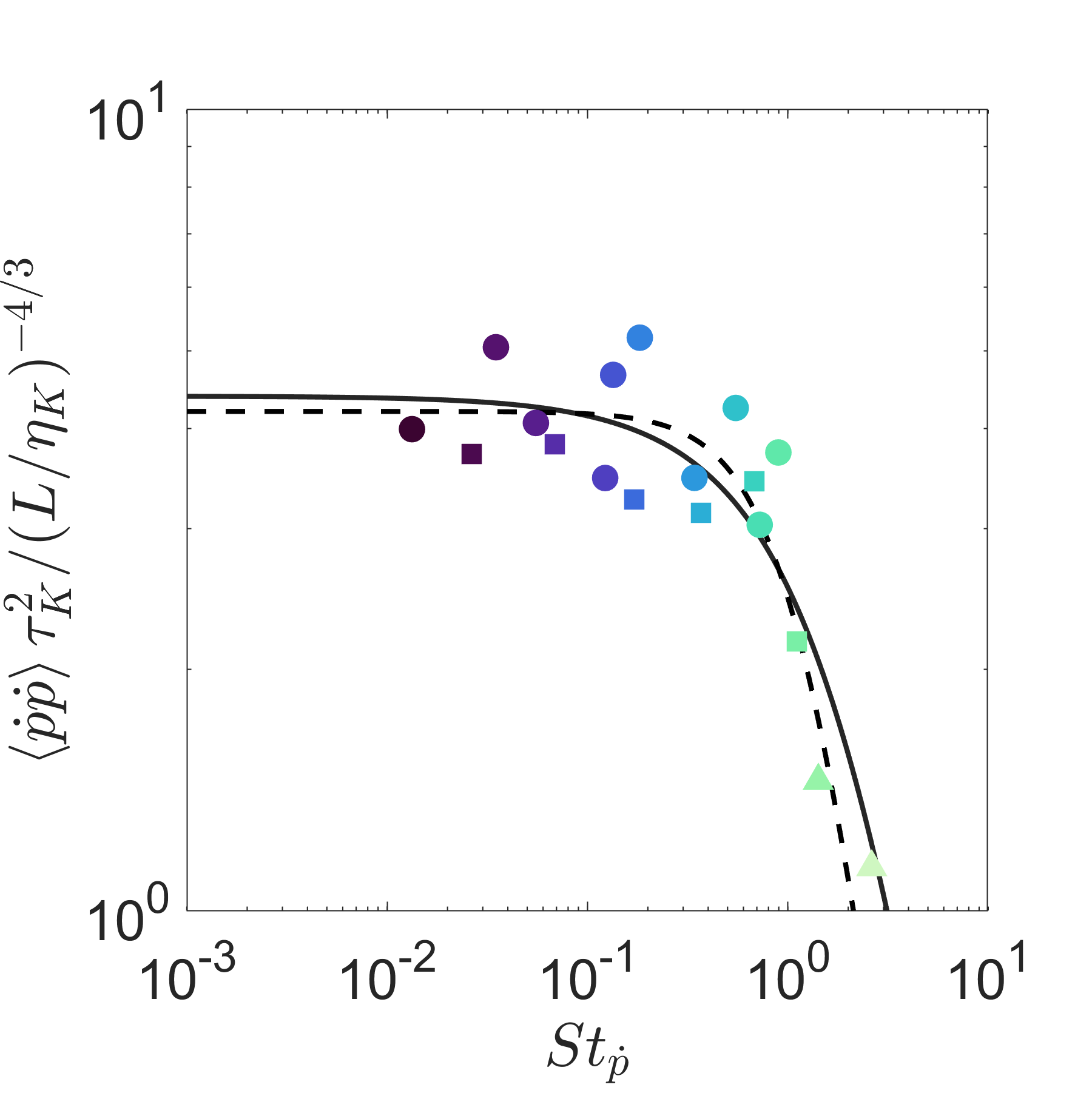} 
         \caption{}
        \label{fig:sigSt}
        \end{subfigure}
        %\vspace{20pt}
    \begin{subfigure}[b]{0.45\textwidth}
        \includegraphics[width=\textwidth]{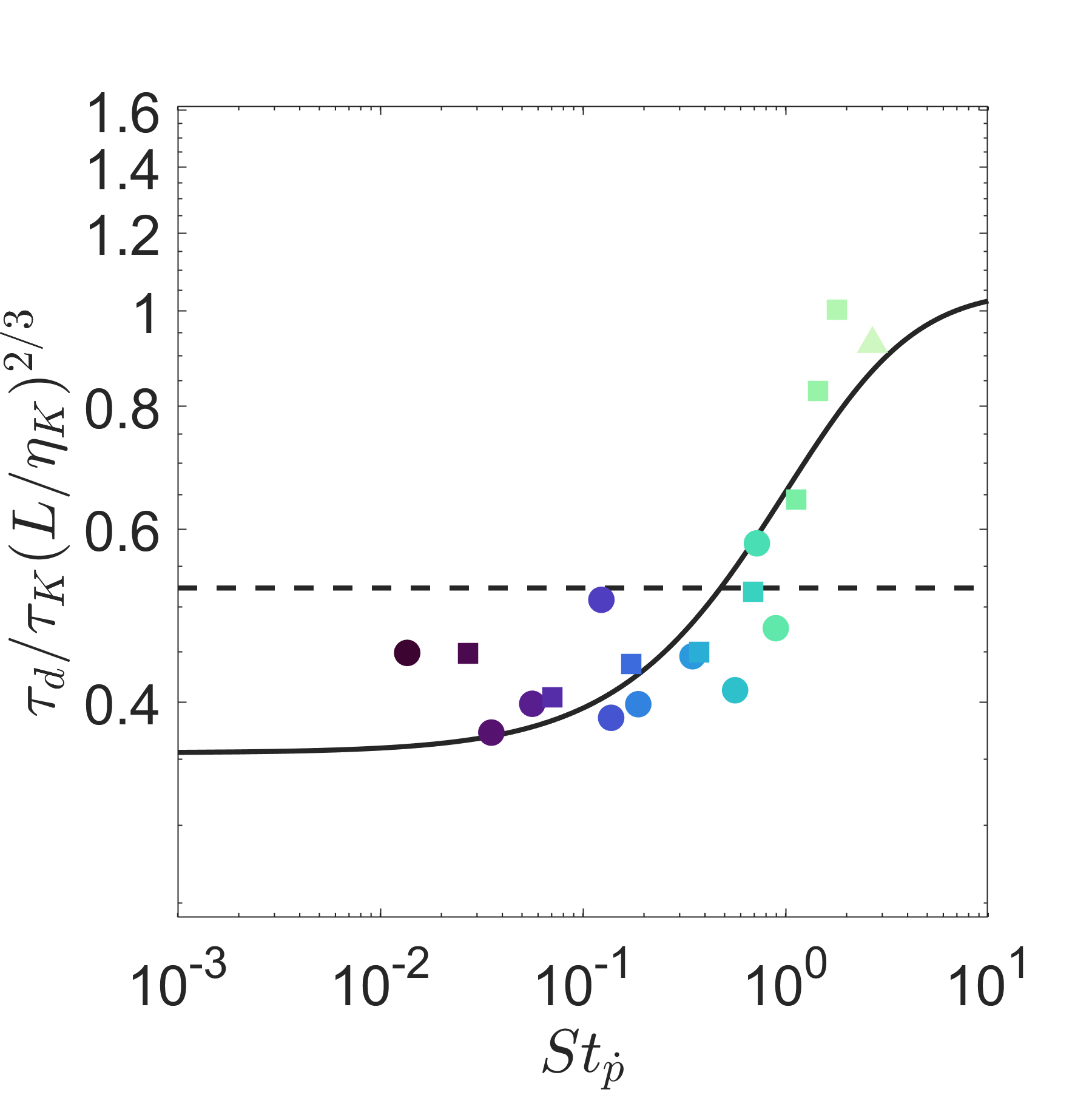} 
         \caption{}
        \label{fig:tauSt}
        \end{subfigure}

 \caption{Evolution of transfer functions of a) variance of tumbling rate and b) correlation time-scale of rotation with respect to rotational Stokes number for fibers of various length and diameter. }
\end{figure}

We test this model on our experimental measurements of $\left<\dot{p}_i\dot{p}_i\right>$ and $\tau_{d1}$ in figures \ref{fig:sigSt} and \ref{fig:tauSt}, respectively.  {We fit equations \ref{eq:04} and \ref{eq:07} simultaneously using two least-squares fits with 3 fitting parameters: the quality factor $Q$, parameter $\alpha$ to rescale the tumbling Stokes number $St_{\dot{p}}=\alpha \omega_L \tau_r$, and parameter $\beta$ to rescale the amplitude of the correlation time. This last parameter is fully justified to compare our prediction with the measurement of the zero-crossing time $\tau_{d1}$ and to compensate the finiteness of the trajectory for the evaluation of the correlation time $\tau_{d2}$.} The best fit {for $\tau_{d1}$} is reached for $Q=0.72$, $\alpha=0.41$, and $\beta=3.12$.  A similar fit to the integral time ($\tau_{d2}$) has yielded the same $Q$ and $\alpha$, but a smaller scaling factor ($\beta$ = 0.64) because of the lower magnitude in $\tau_{d2}$ seen previously (see figure \ref{fig:corr_time}). The dashed lines in each plot represent the predictions from \citet{Bounoua2018}. Results  show that although the previous model is able to predict the variance of tumbling rate,  it fails {to estimate} the evolution of the correlation time. The current model predicts both quantities very well. {Our fitted results for $Q$ shows that the inertial effects begin for $St_{\dot{p}}$ between 0.1 to 1. This agrees qualitatively with the critical $St_{\dot{p}} =0.7$ that we chose by eye in figure \ref{fig:kolm_corr}.} 

\section{Summary and closing comments}
\label{sc:conc}
We report experimental measurements of Lagrangian autocorrelation of tumbling rate of inertial fibers in homogeneous isotropic turbulence. Based on the zero-crossing and the integral of the mean autocorrelation function, we compute two correlation times ($\tau_{d1}$ and $\tau_{d2}$) for fibers with a wide range of length and aspect ratio. The inertia of a fiber is quantified using a rotational Stokes number ($St_{\dot{p}}$) that takes into account fiber length, diameter, as well as the relative density. For low $St_{\dot{p}}$, both correlation times from our measurement follow Kolmogorov's inertial range scaling of $(L/\eta_K)^{2/3}$. This scaling is further supported by the numerically computed zero-crossing times for long slender fibers ($L>10\eta_K$) in \citet{MANSOO2005}. For fibers with high $St_{\dot{p}}$, the rotation rate escapes this prediction. We find that our previous model \citep{Bounoua2018} designed for the variance of tumbling rate does not capture the effect of fiber inertia on correlation time. Instead of assuming the spectrum of background excitation to be a simple Dirac function peaked at the fiber length, we model the excitation as a white noise filtered by a bandpass filter (viscous torque). The evolution of the tumbling rate can then  be described by a Langevin equation with a response time given by the fiber inertia. We show that this model recovers the evolution of both the variance and the correlation time of tumbling rate for the range of explored Stokes number.

{
Recent studies on settling of anisotropic particles in turbulent flows~\citet{Lopez2017,Gustavsson2019,Roy_2019} argue that nonlinear torque plays an important role on the orientation dynamics. It can be surprising that our model, which only considers viscous torque, is in very good agreement with our measurements.  We attribute this contradiction to two main differences between ours and these studies on settling: a) the settling speed in our case is negligible compared to turbulent fluctuations, and b) the size of particles considered in the latter case are generally smaller than the Kolmogorov length. An interesting future work would then be to test this model for fibers whose settling speed is of the same order or larger than the turbulent fluctuations. }

Finally, to fully characterize the rotation of an anisotropic inertial particle, it is necessary to investigate also its spinning motion, which is currently underway in our laboratory. This quantity has been shown to be larger than the tumbling for fibers smaller than the Kolmogorov length due to a preferential alignment \citep{Parsa2012}. Our preliminary results indicate an opposite trend for fibers larger than the Kolmogorov length. Measuring spinning along with tumbling will also help estimating the {total torque and lift experienced by a fiber.}

% comment of inertial effect on the torque/settling, theoretical justification of the expression of the torque
% comment on the spinning

\section{Acknowledgment} 
This work was carried out in the framework of FlexFiT
Project (ANR-17-CE30-0005-01) funded by the French
National Research Agency (ANR), the Labex MEC
Project (No. ANR-10-LABX-0092) and of the A*MIDEX
Project (No. ANR-11-IDEX-0001-02), funded by the
Investissements d’Avenir French Government program
managed by the French National Research Agency (ANR).
%\clearpage
\bibliographystyle{jfm}
\bibliography{PRL2019}

\begin{thebibliography}{25}
\expandafter\ifx\csname natexlab\endcsname\relax\def\natexlab#1{#1}\fi
\def\au#1{#1} \def\ed#1{#1} \def\yr#1{#1}\def\at#1{#1}\def\jt#1{\textit{#1}}
  \def\bt#1{#1}\def\bvol#1{\textbf{#1}} \def\vol#1{#1} \def\pg#1{#1}
  \def\publ#1{#1}\def\arxiv#1{#1}\def\org#1{#1}\def\st#1{\textit{#1}}

\bibitem[Bordoloi \& Variano(2017)]{Bordoloi2017}
{\sc \au{Bordoloi, Ankur~D.} \& \au{Variano, Evan}} \yr{2017}  \at{Rotational
  kinematics of large cylindrical particles in turbulence}.  \jt{Journal of
  Fluid Mechanics}  \bvol{815},  \pg{199--222}.

\bibitem[Bounoua {\em et~al.\/}(2018)Bounoua, Bouchet \& Verhille]{Bounoua2018}
{\sc \au{Bounoua, S.}, \au{Bouchet, G.} \& \au{Verhille, G.}} \yr{2018}
  \at{Tumbling of inertial fibers in turbulence}.  \jt{Physical Review Letters}
   \bvol{121},  \pg{124502}.

\bibitem[Gustavsson {\em et~al.\/}(2019)Gustavsson, Sheikh, Lopez, Naso, Pumir
  \& Mehlig]{Gustavsson2019}
{\sc \au{Gustavsson, K.}, \au{Sheikh, M.~Z.}, \au{Lopez, D.}, \au{Naso, A.},
  \au{Pumir, A.} \& \au{Mehlig, B.}} \yr{2019}  \at{Effect of fluid inertia on
  the orientation of a small prolate spheroid settling in turbulence}.  \jt{New
  J. Phys.}  \bvol{21},  \pg{083008}.

\bibitem[Klein {\em et~al.\/}(2013)Klein, Gibert, B\'{e}rut \&
  Bodenschatz]{Klein2013}
{\sc \au{Klein, S.}, \au{Gibert, M.}, \au{B\'{e}rut, A.} \& \au{Bodenschatz,
  E.}} \yr{2013}  \at{Simultaneous 3d measurement of the translation and
  rotation of finite size particles and the flow field in a fully developed
  turbulent water flow}.  \jt{Meas. Sci. Techol.}  \bvol{24},  \pg{024006}.

\bibitem[Kuperman {\em et~al.\/}(2019)Kuperman, Sabban \& van
  Hout]{Kuperman2019}
{\sc \au{Kuperman, Sofia}, \au{Sabban, Lilach} \& \au{van Hout, Ren{\'{e}}}}
  \yr{2019}  \at{Inertial effects on the dynamics of rigid heavy fibers in
  isotropic turbulence}.  \jt{Physical Review Fluids}  \bvol{4}~(6).

\bibitem[Lopez \& Guazzelli(2017)]{Lopez2017}
{\sc \au{Lopez, Diego} \& \au{Guazzelli, Elisabeth}} \yr{2017}  \at{Inertial
  effects on fibers settling in a vortical flow}.  \jt{Physical Review Fluids}
  \bvol{2}~(2).

\bibitem[Lundell {\em et~al.\/}(2011)Lundell, Söderberg \&
  Alfredsson]{Lundell2011}
{\sc \au{Lundell, Fredrik}, \au{Söderberg, L.~Daniel} \& \au{Alfredsson,
  P.~Henrik}} \yr{2011}  \at{Fluid mechanics of papermaking}.  \jt{Annual
  Review of Fluid Mechanics}  \bvol{43}~(1),  \pg{195--217}.

\bibitem[Marchioli \& Soldati(2013)]{Marchioli2013}
{\sc \au{Marchioli, Cristian} \& \au{Soldati, Alfredo}} \yr{2013}  \at{Rotation
  statistics of fibers in wall shear turbulence}.  \jt{Acta Mechanica}
  \bvol{224}~(10),  \pg{2311--2329}.

\bibitem[Mathai {\em et~al.\/}(2016)Mathai, Neut, van~der Poel \&
  Sun]{Mathai2016}
{\sc \au{Mathai, V.}, \au{Neut, N. W.~M.}, \au{van~der Poel, E.~P.} \& \au{Sun,
  C.}} \yr{2016}  \at{Translational and rotational dynamics of a large buoyant
  sphere in turbulence}.  \jt{Exp. Fluids}  \bvol{57}~(51).

\bibitem[Mccomb \& Chan(1979)]{MCCOMB1979}
{\sc \au{Mccomb, W.~D.} \& \au{Chan, K. T.~J.}} \yr{1979}  \at{Drag reduction
  in fibre suspensions: transitional behaviour due to fibre degradation}.
  \jt{Nature}  \bvol{280}~(5717),  \pg{45--46}.

\bibitem[Michalec {\em et~al.\/}(2017)Michalec, Fouxon, Souissi \&
  Holzner]{Michalec2017}
{\sc \au{Michalec, Fran{\c{c}}ois-Gaël}, \au{Fouxon, Itzhak}, \au{Souissi,
  Sami} \& \au{Holzner, Markus}} \yr{2017}  \at{Zooplankton can actively adjust
  their motility to turbulent flow}.  \jt{Proceedings of the National Academy
  of Sciences}  \bvol{114}~(52),  \pg{E11199--E11207}.

\bibitem[Mordant {\em et~al.\/}(2004)Mordant, Crawford \&
  Bodenschatz]{Mordant2004}
{\sc \au{Mordant, N.}, \au{Crawford, A.M.} \& \au{Bodenschatz, E.}} \yr{2004}
  \at{Experimental lagrangian acceleration probability density function
  measurement}.  \jt{Physica D: Nonlinear Phenomena}  \bvol{193}~(1-4),
  \pg{245--251}.

\bibitem[Parsa {\em et~al.\/}(2012)Parsa, Calvazarini, Toschi \&
  Voth]{Parsa2012}
{\sc \au{Parsa, S.}, \au{Calvazarini, E.}, \au{Toschi, F.} \& \au{Voth, G.~A.}}
  \yr{2012}  \at{Rotation rate of rods in turbulent fluid flow}.  \jt{Phys.
  Rev. Lett.}  \bvol{109},  \pg{134501}.

\bibitem[Parsa \& Voth(2014)]{Parsa2014}
{\sc \au{Parsa, Shima} \& \au{Voth, Greg~A.}} \yr{2014}  \at{Inertial range
  scaling in rotations of long rods in turbulence}.  \jt{Physical Review
  Letters}  \bvol{112}~(2).

\bibitem[Pujara {\em et~al.\/}(2018)Pujara, Oehmke, Bordoloi \&
  Variano]{Pujara2018}
{\sc \au{Pujara, Nimish}, \au{Oehmke, Theresa~B.}, \au{Bordoloi, Ankur~D.} \&
  \au{Variano, Evan~A.}} \yr{2018}  \at{Rotations of large inertial cubes,
  cuboids, cones, and cylinders in turbulence}.  \jt{Physical Review Fluids}
  \bvol{3}~(5).

\bibitem[Pujara \& Variano(2017)]{Pujara2017}
{\sc \au{Pujara, Nimish} \& \au{Variano, Evan~A.}} \yr{2017}  \at{Rotations of
  small, inertialess triaxial ellipsoids in isotropic turbulence}.  \jt{Journal
  of Fluid Mechanics}  \bvol{821},  \pg{517--538}.

\bibitem[Pujara {\em et~al.\/}(2019)Pujara, Voth \& Variano]{Pujara2019}
{\sc \au{Pujara, N.}, \au{Voth, G.~A.} \& \au{Variano, E.}} \yr{2019}
  \at{Scale-dependent alignment, tumbling and stretching of slender rods in
  isotropic turbulence}.  \jt{J. Fluid Mech.}  \bvol{860},  \pg{465--486}.

\bibitem[Roy {\em et~al.\/}(2019)Roy, Hamati, Tierney, Koch \& Voth]{Roy_2019}
{\sc \au{Roy, Anubhab}, \au{Hamati, Rami~J.}, \au{Tierney, Lydia}, \au{Koch,
  Donald~L.} \& \au{Voth, Greg~A.}} \yr{2019}  \at{Inertial torques and a
  symmetry breaking orientational transition in the sedimentation of slender
  fibres}.  \jt{Journal of Fluid Mechanics}  \bvol{875},  \pg{576--596}.

\bibitem[Sabban {\em et~al.\/}(2017)Sabban, Cohen \& van Hout]{Sabban2017}
{\sc \au{Sabban, L.}, \au{Cohen, A.} \& \au{van Hout, R.}} \yr{2017}
  \at{Temporally resolved measurements of heavy, rigid fibre translation and
  rotation in nearly homogeneous isotropic turbulence}.  \jt{Journal of Fluid
  Mechanics}  \bvol{814},  \pg{42--68}.

\bibitem[Sharma(1980)]{Sharma1980}
{\sc \au{Sharma, R.~S.}} \yr{1980}  \at{Drag reduction by fibers}.  \jt{The
  Canadian Journal of Chemical Engineering}  \bvol{58}~(6),  \pg{3--13}.

\bibitem[Shin \& Koch(2005)]{MANSOO2005}
{\sc \au{Shin, Mansoo} \& \au{Koch, Donald}} \yr{2005}  \at{Rotational and
  translational dispersion of fibres in isotropic turbulent flows}.
  \jt{Journal of Fluid Mechanics}  \bvol{540}~(-1),  \pg{143}.

\bibitem[Volk {\em et~al.\/}(2007)Volk, Mordant, Verhille \& Pinton]{Volk2007}
{\sc \au{Volk, R.}, \au{Mordant, N.}, \au{Verhille, G.} \& \au{Pinton, J.-F.}}
  \yr{2007}  \at{Laser doppler measurement of inertial particle and bubble
  accelerations in turbulence}.  \jt{{EPL} (Europhysics Letters)}
  \bvol{81}~(3),  \pg{34002}.

\bibitem[Voth \& Soldati(2017)]{Voth2017}
{\sc \au{Voth, G.} \& \au{Soldati, A.}} \yr{2017}  \at{Anisotropic particles in
  turbulence}.  \jt{Annual Review of Fluid Mechanics}  \bvol{49}~(249).

\bibitem[Xu \& Chen(2013)]{Xu2013}
{\sc \au{Xu, Duo} \& \au{Chen, Jun}} \yr{2013}  \at{Accurate estimate of
  turbulent dissipation rate using {PIV} data}.  \jt{Experimental Thermal and
  Fluid Science}  \bvol{44},  \pg{662--672}.

\bibitem[Zimmermann {\em et~al.\/}(2011)Zimmermann, Gasteuil, Bourgoin, R.,
  Pumir \& Pinton]{Zimmermann2011}
{\sc \au{Zimmermann, R.}, \au{Gasteuil, Y.}, \au{Bourgoin, M.}, \au{R., Bolk},
  \au{Pumir, A.} \& \au{Pinton, J.~F.}} \yr{2011}  \at{Rotational intermittency
  and turbulence induced lift experienced by large particles in a turbulent
  flow}.  \jt{Phys, Rev. Lett.}  \bvol{106},  \pg{154501}.

\end{thebibliography}

%\bibliographystyle{plainnat} 
%% Note the spaces between the initials

\end{document}